\documentclass[aps,pra,reprint,amsmath,showpacs,superscriptaddress]{revtex4-1}

\usepackage{graphicx}
\usepackage{amssymb}
\usepackage{pifont}
\usepackage{color}
\usepackage[mathscr]{euscript}

\def\mathbi#1{\textbf{\em #1}}
\def\hf{\frac{1}{2}}
\def\th{\frac{3}{2}}
\def\fh{\frac{5}{2}}
\def\shf{{\scriptstyle\frac{1}{2}}}
\def\sth{{\scriptstyle\frac{3}{2}}}
\def\sfh{{\scriptstyle\frac{5}{2}}}

\begin{document}

\title{Bose-Einstein condensation with a finite number of particles in a power-law trap}

\author{A. Jaouadi}
\affiliation{Universit\'{e} Paris-Sud, Institut des Sciences Mol\'{e}culaires d'Orsay (ISMO), 91405 Orsay cedex, France}
\affiliation{CNRS, 91405 Orsay cedex, France.}
\affiliation{LSAMA, Department of Physics, Faculty of Science of Tunis, University of Tunis El Manar, 2092  Tunis, Tunisia}

\author{M. Telmini}
\affiliation{LSAMA, Department of Physics, Faculty of Science of Tunis, University of Tunis El Manar, 2092  Tunis, Tunisia}
\affiliation{National Centre for Nuclear Science and Technology, Sidi Thabet Technopark, 2020 Ariana, Tunisia}

\author{E. Charron}
\affiliation{Universit\'{e} Paris-Sud, Institut des Sciences Mol\'{e}culaires d'Orsay (ISMO), 91405 Orsay cedex, France}
\affiliation{CNRS, 91405 Orsay cedex, France.}

\date{\today}

\begin{abstract}
Bose-Einstein condensation (BEC) of an ideal gas is investigated, beyond the thermodynamic limit, for a finite number $N$ of particles trapped in a generic three-dimensional power-law potential. We derive an analytical expression for the condensation temperature $T_c$ in terms of a power series in $x_0=\varepsilon_0/k_BT_c$, where $\varepsilon_0$ denotes the zero-point energy of the trapping potential. This expression, which applies in cartesian, cylindrical and spherical power-law traps, is given analytically at infinite order. It is also given numerically for specific potential shapes as an expansion in powers of $x_0$ up to the second order. We show that, for a harmonic trap, the well known first order shift of the critical temperature $\Delta T_c/T_c \propto N^{-1/3}$ is inaccurate when $N \leqslant 10^{5}$, the next order (proportional to $N^{-1/2}$) being significant. We also show that finite size effects on the condensation temperature cancel out in a cubic trapping potential, \textit{e.g.} $V(\mathbi{r}) \propto r^3$. Finally, we show that in a generic power-law potential of higher order, \textit{e.g.} $V(\mathbi{r}) \propto r^\alpha$ with $\alpha > 3$, the shift of the critical temperature becomes positive. This effect provides a large increase of $T_c$ for relatively small atom numbers. For instance, an increase of about +40\% is expected with $10^4$ atoms in a $V(\mathbi{r}) \propto r^{12}$ trapping potential.
\end{abstract}
\maketitle

\section{Introduction}

Bose-Einstein condensation, first predicted in the years 1924-25 \cite{Bose_1924, Einstein_1924_1925}, was observed in 1995 on rubidium and sodium vapors \cite{Anderson_1995, Davis_1995}. These remarkable experiments have generated since then a clear interest in the critical properties of this dilute macroscopic quantum state of matter. In particular, the transition temperature $T_c$ is a crucial parameter whose exact determination has been a matter of discussion for decades \cite{Dalfovo_1999, Andersen_2004}.

It is indeed rather difficult to estimate the influence of atomic interactions on the condensation temperature for a \textit{uniform} dilute weakly interacting Bose gas. This comes from the fact that this is a many-body problem affected by long-range critical fluctuations which have to be described non-perturbatively \cite{Dalfovo_1999, Andersen_2004, Yukalov_2004}. As a consequence, many different predictions were produced since the late 1950's, like an increase or even a decrease of $T_c$ proportional to $\sqrt{a_s}$, to $a_s$, to $a_s^{3/2}$ or even to $a_s \ln a_s$, where $a_s$ denotes the atomic $s$-wave scattering length \cite{Andersen_2004}. It is only in 1999 that this question was settled with the rigorous demonstration of the linear increase of the critical temperature with $a_s$ \cite{Baym_1999}. It was later shown \cite{Baym_2001} that the relative correction $\Delta T_c / T_c^0$ behaves at second order as $c_1\delta+(c_2'\ln\delta+c_2'')\delta^2$, where $T_c^0$ denotes the condensation temperature of the ideal gas in the thermodynamic limit, and $\delta=\rho^{1/3}a_s$, $\rho$ being the (uniform) atomic density. In this expression, $c_1 \approx 1.32$, $c_2' \approx 19.75$ and $c_2'' \approx 75.7$ \cite{Yukalov_2004}.

On the other hand, in a harmonic trap long-range critical fluctuations are suppressed, and the leading (linear) order correction to $T_c$ can be calculated by simple perturbative methods \cite{Giorgini_1996}. Using a non-perturbative approach, it was shown in 2001 that in such a potential the relative correction $\Delta T_c / T_c^0$ behaves also at second order as $c_1\delta' + (c_2'\ln\delta' + c_2'')\delta'^2$, with $\delta'=a_s/\lambda_T$, $\lambda_T$ being the thermal de Broglie wavelength \cite{Arnold_2001}. In this case, $c_1 \approx -3.426$, $c_2' \approx -45.86$ and $c_2'' \approx -155.0$ \cite{Yukalov_2004}.

These results were then extended to power-law potentials by Zobay \textit{et al} in a series of beautiful papers first using mean-field theory \cite{Zobay_2004a}, then using renormalization group theory \cite{Zobay_2004b, Zobay_2004c}, and later on using variational perturbation theory \cite{Zobay_2005}. These different theoretical studies helped bridging the gap separating \textit{homogeneous} from \textit{inhomogeneous} potentials by showing how an increasingly inhomogeneous potential suppresses the influence of critical fluctuations. In these studies, the shape of the potential could be summarized by a simple parameter $\eta$ which varies between $\hf$ for the uniform gas (homogeneous case) and 2 for the (inhomogeneous) harmonic trap. In Ref. \cite{Zobay_2005}, it was shown for instance that critical fluctuations have a marginal impact on the transition temperature when $\eta > 0.7$ for $\lambda_T \gg a_s$.

The amplitude of the corrections due to atomic interactions increases of course with the number of particles. On the other side, independently of atomic interactions, with a finite number of particles the transition temperature is different from the one obtained in the thermodynamic limit. In the case of a harmonic trap, this finite size effect results in a downward shift of the transition temperature, which is, at lowest order, linear in the zero point energy of the trapping potential \cite{Grossman_1995a, Grossman_1995b, Ketterle_1996}.

\begin{table*}[!t]
\caption{Trap parameters $q$ and $n_i$, with $i \in \{1,\ldots,q\}$, and associated trapping potential $U(\mathbf{r})$. The lines (a), (b) and (c) correspond to cartesian, cylindrical and spherical power-law traps, respectively.}
\begin{ruledtabular}
\begin{tabular}{lcccccc}
                       & $q$ & $n_1$ & $n_2$ & $n_3$ & $U(\mathbf{r})$ \\
  (a) cartesian trap   &  3  &   1   &   1   &   1   & $U_1\!\left|x/d_1\right|^{\alpha_1}\!+
                                                        U_2\!\left|y/d_2\right|^{\alpha_2}\!+
                                                        U_3\!\left|z/d_3\right|^{\alpha_3}$ \\
  (b) cylindrical trap &  2  &   2   &   1   &   -   & $U_1 \left(\rho/d_1\right)^{\alpha_1}\!+
                                                        U_2 \left|z/d_2\right|^{\alpha_2}$ \\
  (c) spherical trap   &  1  &   3   &   -   &   -   & $U_1 \left(r/d_1\right)^{\alpha_1}$ \\
\end{tabular}
\end{ruledtabular}
\label{Tab:traps}
\end{table*}

Following these developments, the present paper deals with higher order finite size corrections in a harmonic trap and in a generic power-law potential. In our approach, the trapped Bose gas is described within the local density approximation. We also adopt a perturbative approach which is only valid in the case of an inhomogeneous trap. Indeed, a non-perturbative treatment is necessary for an accurate estimation of $T_c$ in the case of the uniform gas \cite{Davis_2006}.

The plan of the paper is as follows. In section \ref{Sec:TM}, we introduce our theoretical model. We derive a general expression for finite size effects in section \ref{Sec:FANC}. The next section \ref{Sec:ATBDLGOT} is devoted to the discussion of a specific application in the case of atoms confined in a crossed blue-detuned optical trap. In the final section \ref{Sec:C}, we present a short summary of our main results.

\section{Theoretical Model}
\label{Sec:TM}

\subsection{Generic trapping potential}

For the sake of universality, and in order to treat cartesian, cylindrical, and spherical traps on an equal footing, we follow the approach chosen in Ref. \cite{Yan_2001}. We thus consider a degenerate Bose gas trapped in a generic 3-dimensional power-law potential
\begin{equation}
U(\mathbf{r}) = \sum_{i=1}^{q} U_i\,\left|\frac{r_i}{d_i}\right|^{\alpha_i}\,,
\label{Eq:Generic_Potential}
\end{equation}
where $r_i$, with $i \in \{1,\ldots,q\}$, are the $q$ radial coordinates in the $n_i$-dimensional subspace of the 3-dimensional space (see Table \ref{Tab:traps} for details). $U_i$ and $d_i$ are energy and length scales associated with the trap. The sub-dimensions $n_i$ obviously verify
\begin{equation}
\sum_{i=1}^{q} n_i = 3\,.
\label{Eq:Dimension}
\end{equation}
Table \ref{Tab:traps} gives, with the practical expression of the trapping potential $U(\mathbf{r})$ of Eq. (\ref{Eq:Generic_Potential}), a summary of the different parameters involved in cartesian, cylindrical and spherical coordinate systems.

In order to study finite size corrections to the ideal condensation temperature $T_c^0$ obtained in the thermodynamic limit, we first define the volume of the trap $V(\varepsilon)$ which includes the portion of space where $U(\mathbf{r}) \leqslant \varepsilon$. The variable $\varepsilon$ denotes here an arbitrary energy. Formally, this volume can be written as
\begin{equation}
V(\varepsilon) = \iiint_{U(\mathbf{r}) \leqslant \varepsilon} d^3r\,.
\label{Eq:pseudo-volume-def}
\end{equation}
For an isotropic \textit{harmonic} trap of angular frequency $\omega$, this definition gives a volume of $\frac{4}{3} \pi a_{h}^3$ at the energy $\varepsilon = \hf\hbar\omega$, where $a_{h}$ is the usual characteristic length of the harmonic trap. In the general case, integration over the 3-dimensional space yields the expression
\begin{equation}
V(\varepsilon) = \frac{\theta\,\widetilde{\Pi}}{\Gamma(\eta+\shf)}\,\varepsilon^{\eta-\frac{1}{2}}\,,
\label{Eq:pseudo-volume}
\end{equation}
where
\begin{equation}
\widetilde{\Pi} = \prod_{i=1}^{q} \Gamma\Big(1+\frac{n_i}{\alpha_i}\Big)\,d_i^{\,n_i}\,U_i^{-\frac{n_i}{\alpha_i}}\,,
\label{Eq:Pi}
\end{equation}
and where $\theta = 8$, $2\pi$ or $4\pi/3$ in the case of a cartesian, cylindrical or spherical trap respectively. In these equations, $\Gamma(x)$ denotes the complete gamma function \cite{Abramowitz_1964}. In addition, in Eq. (\ref{Eq:pseudo-volume}), $\eta$ denotes a very important parameter which characterizes the shape of the potential. This parameter is defined by
\begin{equation}
\eta = \frac{1}{2} + \sum_{i=1}^{q} \frac{n_i}{\alpha_i}\,.
\label{Eq:eta}
\end{equation}

One can see in this last equation that the shape parameter $\eta$ equals 2 in a harmonic trap corresponding to $\alpha_i = 2~(\forall i)$, while $\eta$ equals $\frac{1}{2}$ for a square-well potential corresponding to $\alpha_i \rightarrow \infty~(\forall i)$. Intermediate values $\frac{1}{2} < \eta < 2$ correspond to the different cases of power-law potentials, as defined in Eq. (\ref{Eq:Generic_Potential}).

\subsection{Density of states}

\begin{figure}[!t]
\centering
\includegraphics*[width=8.6cm,clip=true]{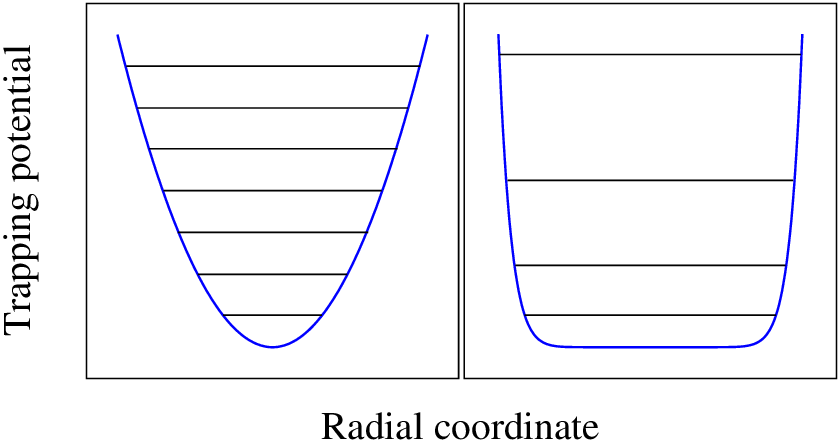}
\caption{(Color online) Schematic representation of a one-dimensional harmonic trapping potential $V(x) \propto x^2$ (solid blue line in the left panel) and of a one-dimensional power-low trapping potential $V(x) \propto x^{12}$ (solid blue line in the right panel), with their associated energy level structures.}
\label{Fig:Pot}
\end{figure}

In the theoretical description of Bose-Einstein condensation within the local density approximation, three-dimensional integrals of functions of the trapping potential $U(\mathbf{r})$ are common. They can be greatly simplified using the following transformation \cite{Zobay_2004a}
\begin{equation}
\iiint f\big(U(\mathbf{r})\big) d^3r = \int f(\varepsilon) v(\varepsilon) d\varepsilon\,,
\label{Eq:Transf}
\end{equation}
where the spatial volume $v(\varepsilon)d\varepsilon$ of the equi-potential shell of $U(\mathbf{r})$ with width $d\varepsilon$ at energy $\varepsilon$ can be obtained from Eq. (\ref{Eq:pseudo-volume}). This yields
\begin{equation}
v(\varepsilon) = \frac{\theta\,\widetilde{\Pi}}{\Gamma(\eta-\frac{1}{2})}\,\varepsilon^{\eta-\frac{3}{2}}\,.
\label{Eq:vE}
\end{equation}
The transformation (\ref{Eq:Transf}) is for instance helpful for the determination of the density of states $g(\varepsilon)$ associated with the generic trapping potential $U(\mathbf{r})$. In a semi-classical approach, this density is given by
\begin{equation}
g(\varepsilon) = \frac{1}{h^3} \iiint d^3\mathbf{r} \iiint d^3\mathbf{p}\; \delta\big(\varepsilon-\varepsilon_{\mathrm{cl}}(\mathbf{r},\mathbf{p})\big)\,,
\label{Eq:gE-def}
\end{equation}
where $\mathbf{p}$ is the momentum vector of the atom and where, in the case of the ideal gas, $\varepsilon_{\mathrm{cl}}(\mathbf{r},\mathbf{p})$ denotes the dispersion relation
\begin{equation}
\varepsilon_{\mathrm{cl}}(\mathbf{r},\mathbf{p}) = \frac{\mathbf{p}^2}{2m} + U(\mathbf{r})\,,
\end{equation}
$m$ being the atomic mass. In the expression (\ref{Eq:gE-def}), $\delta(x)$ denotes the Dirac delta function. Since $\varepsilon_{\mathrm{cl}}(\mathbf{r},\mathbf{p})$ does not depend on the orientation of the momentum vector but on its magnitude only, one can integrate over the momentum coordinate to find \cite{Davis_2001}
\begin{equation}
g(\varepsilon) = \frac{m^{\frac{3}{2}}}{2^{\frac{1}{2}}\pi^2\hbar^3} \iiint \big[\varepsilon-U(\mathbf{r})\big]^{\frac{1}{2}} d^3\mathbf{r}\,.
\end{equation}
Integrating this expression using the transformation (\ref{Eq:Transf}) then yields
\begin{equation}
g(\varepsilon) = \frac{1}{\hbar^3} \left(\frac{m}{2\pi}\right)^\frac{3}{2} \frac{\theta\,\widetilde{\Pi}}{\Gamma(\eta+1)}\, \varepsilon^\eta\,.
\label{Eq:gE}
\end{equation}
Consequently, $g(\varepsilon)$ is proportional to $\varepsilon^2$ for a harmonic trap and to $\sqrt{\varepsilon}$ in the homogeneous limit corresponding to $\eta \rightarrow \hf$. This strong evolution of the density of states as a function of the shape of the potential could already be inferred from Fig. \ref{Fig:Pot}, which shows a schematic representation of a one dimensional harmonic potential (left side) and of a one-dimensional power-law potential $V(x) \propto x^{12}$ (right side), with their associated energy level structures. These two traps of different shapes are characterized by very different energy level structures, and are therefore characterized by very different densities of states. Since the condensation process consists in an amplified growth of the ground state population resulting from stimulated collisions of atoms in a thermal reservoir, these very different densities of states lead to different condensation temperatures \cite{Jaouadi_2010}, but also, as we will see hereafter, to different corrections due to finite size effects.

\subsection{Ideal condensation temperature in the thermodynamic limit}

Assuming that, at the condensation temperature, the number of condensed atoms is negligible, one can evaluate the total number of particles $N$ as
\begin{equation}
N = \iiint n_T(\mathbf{r})\,d^3\mathbf{r}\,,
\label{Eq:Ntot}
\end{equation}
where the spatial density distribution of the thermal gas is given, in the local density approximation, by
\begin{equation}
n_T(\mathbf{r}) = \frac{1}{\lambda_T^3}\,g_{\th}\!\left(\exp\left[\frac{\mu-U(\mathbf{r})}{k_BT}\right]\right)\,.
\label{Eq:density}
\end{equation}
In this expression $\mu$ is the chemical potential and $\lambda_T$ the thermal de Broglie wavelength at the absolute temperature $T$
\begin{equation}
\lambda_T = \sqrt{\frac{2\pi\hbar^2}{mk_BT}}\,.
\label{Eq:lambdaT}
\end{equation}
$g_s(z)$ denotes the Bose function of order $s$, which, for our purposes, can be considered as defined in terms of its series representation
\begin{equation}
g_s(z) = \sum_{k=1}^{\infty} \frac{z^k}{k^s}\,.
\label{Eq:Bosefunction}
\end{equation}

In the case of an ideal gas in the thermodynamic limit, the critical value of the chemical potential is $\mu=0$. Introducing this value in Eq. (\ref{Eq:density}) and integrating Eq. (\ref{Eq:Ntot}) using the transformation (\ref{Eq:Transf}) yields
\begin{equation}
N = \frac{\theta\,\widetilde{\Pi}}{\hbar^3} \left(\frac{m}{2\pi}\right)^\th \zeta(\eta+1) \Big[k_BT^{0}_{c}\Big]^{\eta+1}\,,
\label{Eq:Tc0}
\end{equation}
from which the critical temperature $T_c^0$ of the ideal gas in the thermodynamic limit can be obtained. $T_c^0$ therefore varies as $N ^{1/(\eta+1)}$, the shape parameter $\eta$ being defined in Eq. (\ref{Eq:eta}). Since $\hf \leqslant \eta \leqslant 2$, we see already that the shape of the potential has a strong impact on the dependence of the critical temperature $T_c^0$ with the number of particles \cite{Jaouadi_2010}.

\section{Finite atom number corrections}
\label{Sec:FANC}

\subsection{Derivation of the shift}

For a finite number of ideal particles, the zero-point energy of the trapping potential can not be neglected anymore and the critical value of the chemical potential is bounded by $\mu = \varepsilon_0$. Following the method used previously to derive Eq. (\ref{Eq:Tc0}), one can introduce this value of $\mu$ in Eq. (\ref{Eq:density}) and integrate Eq. (\ref{Eq:Ntot}) using the transformation (\ref{Eq:Transf}) to obtain a new expression for the total number of atoms which is very similar to Eq. (\ref{Eq:Tc0}):
\begin{equation}
N = \frac{\theta\,\widetilde{\Pi}}{\hbar^3} \left(\frac{m}{2\pi}\right)^\th \frac{\left(\eta|x_0\right)_{x_0}^{\infty}}{\Gamma(\eta-\hf)} \Big[k_BT_{c}\Big]^{\eta+1}\,,
\label{Eq:Tc}
\end{equation}
where
\begin{equation}
\left(\eta|x_0\right)_{a}^{b} = \int_a^b g_{\th}\!\left(e^{x_0-x}\right) x^{\eta-\th}\,dx\,,
\label{Eq:Ietaq}
\end{equation}
and where $x_0 = \varepsilon_0/k_BT_{c}$. Note that the change of variable $x = \varepsilon/k_BT_{c}$ was used. The condensation temperature $T_c = T_c^0 + \Delta T_c$ now includes the corrections induced by the presence of a finite number of particles. Comparing Eqs. (\ref{Eq:Tc0}) and (\ref{Eq:Tc}), we find
\begin{equation}
\left(\frac{T_c}{T_c^0}\right)^{\eta+1} = \frac{\zeta(\eta+1)\Gamma(\eta-\hf)}{\left(\eta|x_0\right)_{x_0}^{\infty}}\,.
\label{Eq:CompTc}
\end{equation}

The calculation of the integral $\left(\eta|x_0\right)_{x_0}^{\infty}$ requires to separate long from short range effects using
\begin{equation}
\left(\eta|x_0\right)_{x_0}^{\infty} = \left(\eta|x_0\right)_{0}^{\infty} - \left(\eta|x_0\right)_{0}^{x_0}\,.
\end{equation}
The first integral is easily obtained using the Bose function (\ref{Eq:Bosefunction}) as
\begin{equation}
\left(\eta|x_0\right)_{0}^{\infty} = \Gamma(\eta-\shf)\,g_{\eta+1}\!\left(e^{x_0}\right)\,,
\label{Eq:long}
\end{equation}
while the second integral can be expressed as
\begin{equation}
\left(\eta|x_0\right)_{0}^{x_0} = x_0^{\eta-\hf} \int_0^{1} g_{\th}\!\left(e^{x_0(1-u)}\right) u^{\eta-\th}\,du\,.
\label{Eq:short}
\end{equation}
In these last two expressions, one can expand the two Bose functions $g_{\eta+1}\!\left(e^{x_0}\right)$ and $g_{\th}\!\left(e^{x_0(1-u)}\right)$ using the series representations of these functions about zero \cite{Gradshteyn_1980}.

For non-integer values of $\eta$ ($\eta \in \mathbb{R}\!\smallsetminus\!\mathbb{N}$), this yields
\begin{equation}
\left(\eta|x_0\right)_{x_0}^{\infty} = A_{\eta}\,x_0^{\eta} + \sum_{k=0}^{\infty} \left[ B_{k,\eta} - C_{k,\eta}\,x_0^{\eta-\hf} \right] x_0^k\,,
\label{Eq:Real}
\end{equation}
where
\begin{subequations}
\begin{eqnarray}
A_{\eta}   & = & \Gamma(\eta-\shf)\cos(\eta\pi)\Gamma(-\eta)\\
B_{k,\eta} & = & \Gamma(\eta-\shf)\zeta(\eta+1-k)/k!\\
C_{k,\eta} & = & \Gamma(\eta-\shf)\zeta(\sth-k)/\Gamma(\eta+k+\shf)\,,
\end{eqnarray}
\label{Eq:ABC}
\end{subequations}
and for $\eta \in \mathbb{N}$, one obtains
\begin{equation}
\left(\eta|x_0\right)_{x_0}^{\infty} = I_{\eta}(x_0)\,x_0^{\eta} + \sum_{k = 0}^{\infty} \left[B_{\substack{k,\eta\\ k\ne \eta}}-C_{k,\eta}\,x_0^{\eta-\hf}\right] x_0^k\,,
\label{Eq:Integer}
\end{equation}
where
\begin{equation}
I_{\eta}(x_0) = \frac{\Gamma(\eta-\hf)}{\eta!} \left[ - \ln(x_0) + \sum_{k = 1}^{\eta} \frac{1}{k} \right]\,,
\end{equation}
the parameters $B_{k,\eta}$ and $C_{k,\eta}$ being already defined in Eqs. (\ref{Eq:ABC}b) and (\ref{Eq:ABC}c).

Practically, for small values of $x_0$, one can truncate the sums in Eqs. (\ref{Eq:Real}) and (\ref{Eq:Integer}) to the desired level of accuracy and introduce the expression obtained for $\left(\eta|x_0\right)_{x_0}^{\infty}$ in Eq. (\ref{Eq:CompTc}) to get the correction $\Delta T_c = T_c - T_c^0$ due to the presence of a finite number of particles.

\subsection{Discussion}

\subsubsection{The harmonic trap}

In the special case of a harmonic trap $(\eta=2)$, and for $x_0 \leqslant 0.1$, the truncation can be limited to second order in $x_0$, and a simple expansion of $T_c/T_c^0$ [Eq. (\ref{Eq:CompTc})] in powers of $x_0$ yields at second order
\begin{equation}
\frac{\Delta T_c}{T_c^0} = - \frac{\zeta(2)}{3\zeta(3)}x_0 + \frac{\zeta(\th)}{3\zeta(3)\Gamma(\fh)}x_0^\th - \alpha(x_0) x_0^2\,,
\label{Eq:Harm}
\end{equation}
where
\begin{equation}
\alpha(x_0) = \frac{1}{3\zeta(3)} \left[ \frac{2\left[\zeta(2)\right]^2}{3\zeta(3)} - \frac{3}{4} + \frac{\ln(x_0)}{2} \right]\,.
\end{equation}
In this expression, one can recognize the well-known first order correction $-(\zeta(2)/[3\zeta(3)])\,x_0 \simeq -0.4561\,x_0$ of the harmonic trap \cite{Pethick_2008}. Note however that the next order correction $+(\zeta(\th)/[3\zeta(3)\Gamma(\fh)])\,x_0^{3/2} \simeq +0.5449\,x_0^{3/2}$ should be taken into account when $x_0 \geqslant 10^{-3}$ for an accurate estimation of the correction to $T_c$. Up to the order $x_0^{3/2}$, the correction (\ref{Eq:Harm}) to the condensation temperature in a harmonic trap can be rewritten as a function of the number of atoms as
\begin{eqnarray}
\frac{\Delta T_c}{T_c^0} & \simeq &
-\frac{\zeta(2)\,\omega_m}{2\zeta(3)^{\frac{2}{3}}\,\overline{\omega}} N^{-\frac{1}{3}}
+ \zeta(\sth) \left(\frac{2\,\omega_m^3}{3\pi\zeta(3)\,\overline{\omega}^3}\right)^{\!\!\frac{1}{2}} N^{-\frac{1}{2}}\nonumber\\
                         & \simeq &
-0.7275\left(\frac{\omega_m}{\overline{\omega}}\right) N^{-\frac{1}{3}}
+ 1.098\left(\frac{\omega_m}{\overline{\omega}}\right)^{\!\!\frac{3}{2}} N^{-\frac{1}{2}}
\label{Eq:HarmN}
\end{eqnarray}
where
\begin{subequations}
\begin{equation}
\omega_m          = \frac{1}{3}\sum_{i=1}^{q} n_i\omega_i
\end{equation}
and
\begin{equation}
\overline{\omega} = \left(\,\prod_{i=1}^{q} \omega_i^{n_i}\right)^{\frac{1}{3}}
\end{equation}
\end{subequations}
denote the arithmetic and geometric means of the oscillator frequencies $\omega_i$ associated with each radial coordinate.

\begin{figure}[!t]
\centering
\includegraphics*[width=8.6cm,clip=true]{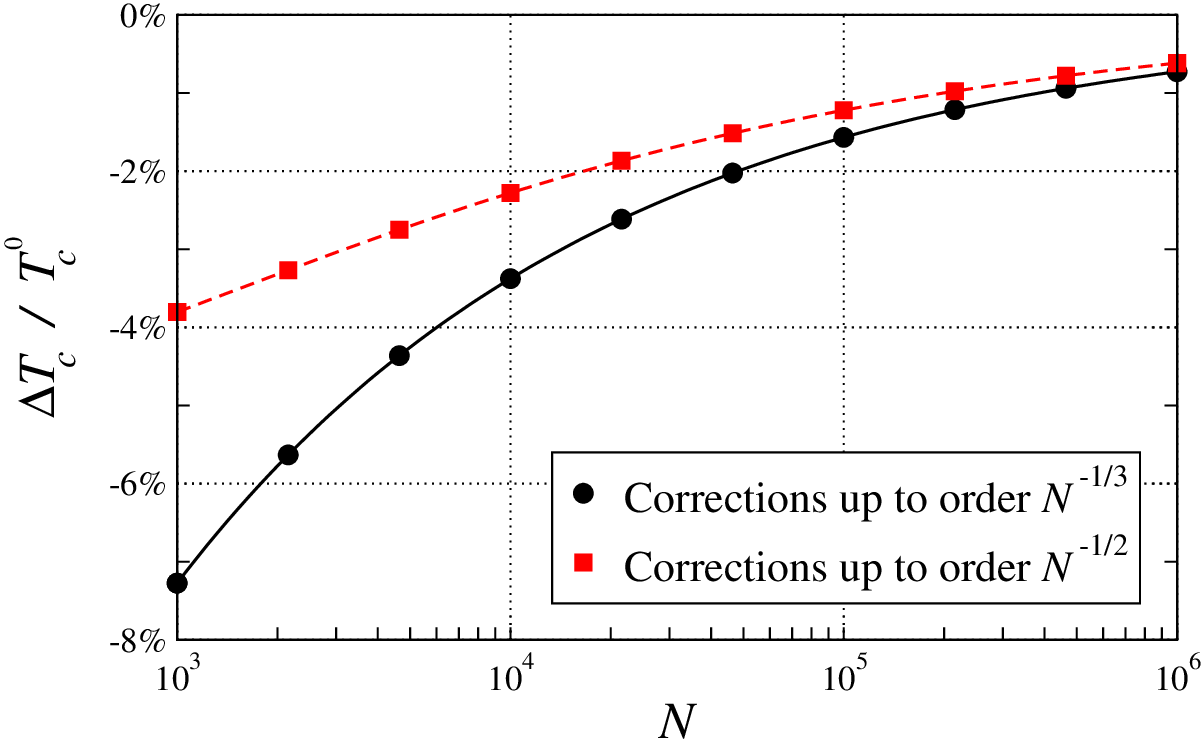}
\caption{(Color online) Relative correction $\Delta T_c / T_c^0$ due to finite size corrections as a function of the number of atoms $N$ (logarithmic scale) in an isotropic harmonic trap. The corrections up to order $N^{-1/3}$ and  $N^{-1/2}$ in Eq. (\ref{Eq:HarmN}) are shown as a solid line with black circles and a dashed line with red squares, respectively.}
\label{Fig:dTc-N}
\end{figure}

This relative correction $\Delta T_c/T_c^0$ in shown in Fig. \ref{Fig:dTc-N} as a function of the number of particles $N$ in the case of an isotropic harmonic trap $(\omega_m=\overline{\omega})$. In this figure, the solid black line (with circles) represents  the relative correction obtained using the first order correction only, whereas the dashed red line (with squares) represents the relative correction obtained using the more accurate expression (\ref{Eq:HarmN}).

One can see in Fig. \ref{Fig:dTc-N} that the well known first order correction $-0.7275\, (\omega_m/\overline{\omega})\, N^{-\frac{1}{3}}$ is inaccurate for small atom numbers. This is due to the fact that, beyond the first order correction, the next order is not a second order correction, which would be proportional to $N^{-\frac{2}{3}}$, but rather a correction proportional to $N^{-\frac{1}{2}}$. When using only the first order correction found in the literature, the error made on the shift of the critical temperature is of about 91\%, 48\%, 29\% and 18\% for $N = 10^3$, $10^4$, $10^5$ and $10^6$, respectively. We can therefore conclude that the last term of Eq. (\ref{Eq:HarmN}), proportional to $N^{-\frac{1}{2}}$, should be included for an accurate estimation of finite size corrections to $T_c$ in a harmonic trap when $N \leqslant 10^{5}$.

\subsubsection{The power-law trap}

Beyond the simple case of the harmonic trap, a general expansion of $T_c/T_c^0$ [Eq. (\ref{Eq:CompTc})] in powers of $x_0$ can be derived for the two cases $\eta \in \mathbb{R}\!\smallsetminus\!\mathbb{N}$ and $\eta \in \mathbb{N}$. We give in Table \ref{Tab:Tc}, at second order, practical expressions for a set of specific values of $\eta$ between 0.75 and 2, corresponding to different kinds of inhomogeneous potentials.

\begin{table}[!t]
\caption{Practical second order expressions of the shift to $T_c$ due to finite size effects for different values of the shape parameter $\eta$. $T_c^0$ denotes the condensation temperature obtained in the thermodynamic limit and $x_0 = \varepsilon_0 / k_BT_c$ denotes the ratio of the zero point energy of the trapping potential to the thermal energy $k_BT_c$.}
\begin{ruledtabular}
\begin{tabular}{cc}
$\eta$ & $\Delta T_c / T_c^0$\\[0.1cm]
\hline\\
           & $  0.8393\,x_0^{1/4} + 0.9685\,x_0^{1/2} + 0.2239\,x_0^{3/4}$\\
3/4        & $+ 0.3037\,x_0       - 0.2536\,x_0^{5/4} + 0.2064\,x_0^{3/2}$\\
           & $- 0.2921\,x_0^{7/4} + 0.3222\,x_0^2 + O[x_0^{9/4}]$\\[0.2cm]
\hline\\[-0.1cm]
           & $  0.8960\,x_0^{1/2} + [ 0.9003 + 0.3040 \ln x_0 ]\,x_0$\\
1          & $+ [ 0.6474 + 0.8171 \ln x_0 ]\,x_0^{3/2}$\\
           & $+ 0.139 [ 0.201 + \ln x_0 ] [ 11.01 + \ln x_0 ]\,x_0^2 + O[x_0^{5/2}]$\\[0.2cm]
\hline\\[-0.1cm]
5/4        & $  0.8652\,x_0^{3/4} - 1.399\,x_0 + 0.8440\,x_0^{5/4}$\\
           & $+ 1.216\,x_0^{3/2} - 4.209\,x_0^{7/4} + 5.675\,x_0^2 + O[x_0]^{9/4}$\\[0.2cm]
\hline\\[-0.1cm]
3/2        & 0 (to all orders in $x_0$, see text for details)\\[0.3cm]
\hline\\[-0.1cm]
7/4        & $-0.5662\,x_0 + 0.6653\,x_0^{5/4} - 0.5636\,x_0^{7/4}$\\
           & $+ 1.098\,x_0^2 + O[x_0^{9/4}]$\\[0.2cm]
\hline\\[-0.1cm]
2          & $- 0.4561\,x_0 + 0.5449\,x_0^{3/2}$\\
           & $+ [ 0.2082 + 0.1387 \ln x_0 ]\,x_0^2 + O[x_0^{5/2}]$\\[0.1cm]
\end{tabular}
\end{ruledtabular}
\label{Tab:Tc}
\end{table}

One can first notice in Table \ref{Tab:Tc} that the case $\eta=3/2$ is a special case. Indeed, for $\eta=3/2$ all finite atom number corrections cancel out in Eq. (\ref{Eq:Real}) except for the constant term $B_{0,\th}=\zeta(\sfh)$. $T_c$ is then identical to $T_c^0$ for all $N$. This cancellation comes from the fact that incidentally $C_{k,3/2} = B_{k+1,3/2}$ and $A_{3/2} = 0$ [see Eqs. (\ref{Eq:ABC})]. This is the only value of the shape parameter $\eta$ where there is rigorously no finite atom number correction, whatever the number of trapped atoms. This special case corresponds for instance to an isotropic cubic trap, with $V(\mathbi{r}) \propto r^3$, but also to a trapping potential of cylindrical symmetry such as $V(\mathbi{r}) \propto \rho^3 + |z|^3$ or to $V(\mathbi{r}) \propto |x|^3 + |y|^3 + |z|^3$ in the case of a cartesian power-law trap. It also applies for instance to $V(\mathbi{r}) \propto \rho^4 + z^2$, or to any combination of power-law potentials for which $\eta =  3/2$ [see Eq. (\ref{Eq:eta})].

\begin{figure}[!t]
\centering
\includegraphics*[width=8.6cm,clip=true]{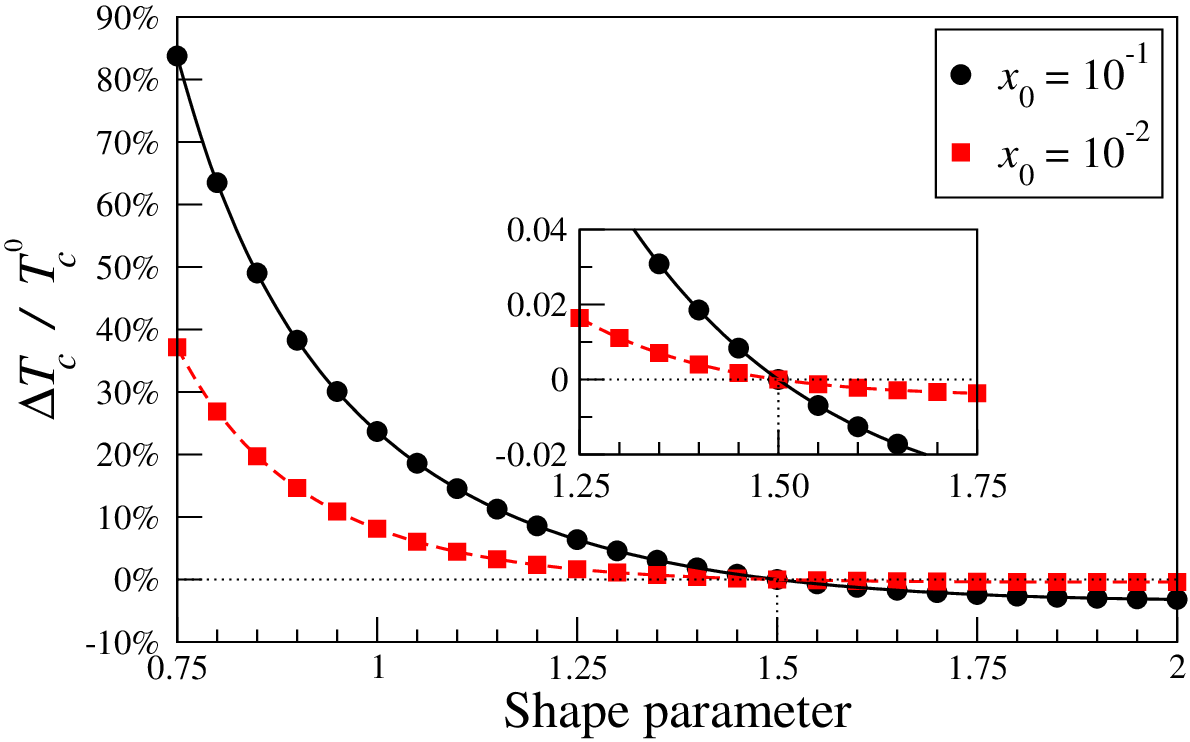}
\caption{(Color online) Relative correction $\Delta T_c / T_c^0$ (in \%) due to finite size effects as a function of the shape parameter $\eta$ for $x_0=10^{-1}$ (solid line with black circles) and $x_0=10^{-2}$ (dashed line with red squares). The inset is a simple magnification of the region surrounding $\eta=3/2$. Note that $\Delta T_c / T_c^0 = 0$ for $\eta=3/2$, $\forall x_0$.}
\label{Fig:dTc-eta}
\end{figure}

The variation of the relative correction $\Delta T_c / T_c^0$ is shown as a function of the shape parameter $\eta$ in Fig. \ref{Fig:dTc-eta} for $x_0=0.1$ (solid line with black circles) and $x_0=0.01$ (dashed line with red squares). From the magnified inset of this figure, one can notice again that the point $\eta=3/2$ plays a particular role since trapping potentials with $\eta > 3/2$ are characterized by a decrease of $T_c$ due to finite size effects, while $\eta < 3/2$ leads, on the contrary, to an increase of $T_c$. This could already be inferred from Table \ref{Tab:Tc}, where it is seen that the lowest order correction is positive when $\eta < 3/2$ and negative when $\eta > 3/2$.

In addition, one can notice in Fig. \ref{Fig:dTc-eta} that, even when $x_0 \ll k_BT_c$, the increase in $T_c$ becomes significantly large when $\eta \leqslant 1$. Indeed, for $\eta = 3/4$, an increase of 37\% is expected when $x_0 = 0.01$, and an increase of 84\% is expected when $x_0 = 0.1$. A large upward shift of $T_c$ can therefore be expected in the case of a generic high-order power-law potential, like $V(\mathbi{r}) \propto r^\alpha$ with $\alpha > 3$ for instance. One should note here that we have limited our study to $0.75 \leqslant \eta \leqslant 2$, \textit{i.e.} to inhomogeneous potentials since a non-perturbative treatment would be necessary for an accurate estimation of $\Delta T_c$ in the homogeneous limit $\eta\rightarrow\hf$.

In order to check the influence of these corrections in a realistic trapping situation, we now turn to the case of a rubidium gas trapped by blue-detuned Laguerre-Gaussian optical beams.\\

\section{Application to blue-detuned Laguerre-Gaussian optical traps}
\label{Sec:ATBDLGOT}

\subsection{Introduction and model system}

Recently, we have proposed an original route to achieve Bose-Einstein condensation using dark power-law laser traps \cite{Jaouadi_2010}. Experimentally, it has been demonstrated that cold atoms can be trapped and guided in such optical setups \cite{Mestre_2010}. These traps are created with two crossing blue-detuned Laguerre-Gaussian optical beams. A simple control of their azimuthal order $\ell$ allows for the exploration of a wide range of situations, with atoms trapped in confining power-law potentials in one, two, and three dimensions. It was shown that, for a fixed atom number, higher transition temperatures are obtained in configurations where $\eta < 2$, compared to a harmonic trap $(\eta = 2)$ of the same size. We investigate here finite size effects in such optical traps.

To create a three-dimensional trap, we consider an all-optical configuration consisting of two perpendicularly crossing Laguerre-Gauss laser modes of same radial index $p=0$ and azimuthal index $\ell$. The polarizations of the two beams are chosen to be orthogonal to avoid any interference pattern. The first beam, circularly symmetric, propagates along the $z$ direction and provides trapping in the $(x,y)$ plane. Trapping in the third dimension is provided by another strongly elongated Laguerre-Gaussian beam shaped elliptically in the form of a light sheet and propagating in the $x$ direction. If the two beams are characterized by the same laser power, waist and detuning, the corresponding potential near the trap center can be described by
\begin{equation}
V(\rho,z) = U_0 \left( \rho^{2\ell} + z^{2\ell} \right)\,.
\label{Eq:Opt-Pot}
\end{equation}
$U_0$ is then a simple function of the laser power, detuning and waist of the two beams \cite{Jaouadi_2010}, and $\eta=(\ell+3)/(2\ell)$.

\begin{figure}[!t]
\centering
\includegraphics*[width=8.6cm,clip=true]{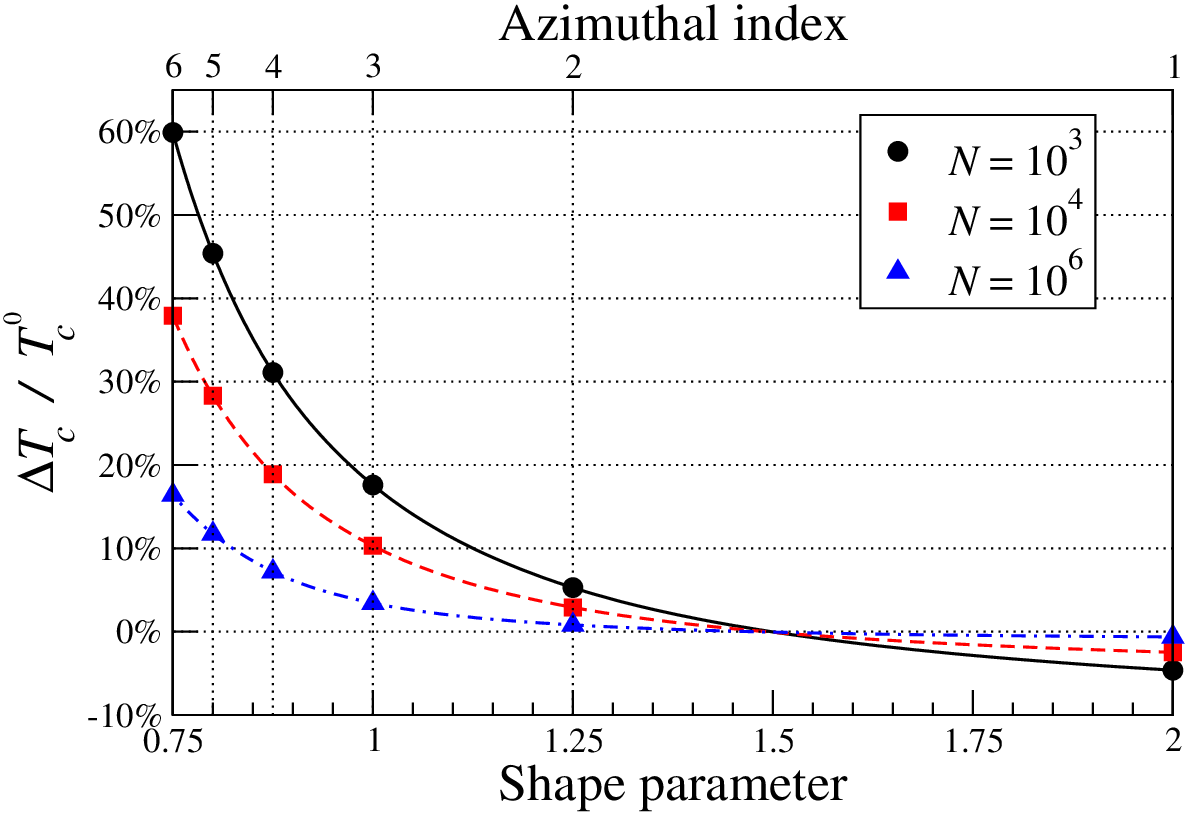}
\caption{(Color online) Relative correction $\Delta T_c / T_c^0$ (in \%) due to finite size effects as a function of the shape parameter $\eta$ (lower scale) and of the azimuthal index $\ell$ (upper scale) for two Laguerre-Gaussian beams of power $P = 5\,$W and detuning $\delta = 10\,$THz. The solid black line with circles corresponds to $N = 10^3$ particles, the red dashed line with squares to $N=10^4$ and the blue dash-dotted line with triangles to $N=10^6$. The volume of the trap has been chosen such that, for each atom number, the condensation temperature $T_c^0$ obtained in the thermodynamic limit for $\eta=2$ is 100\,nK (see \cite{Jaouadi_2010} for details).}
\label{Fig:dTc-l}
\end{figure}

\subsection{Finite size effects}

The relative correction $\Delta T_c / T_c^0$ due to finite size effects in such optical traps is shown in Fig. \ref{Fig:dTc-l} as a function of the shape parameter $\eta$ (lower scale) or equivalently of the azimuthal index $\ell$ (upper scale), for realistic values of all optical parameters (see figure caption). One can notice in this figure that, even for large condensate occupation numbers, the shift of the condensation temperature due to finite size effects is not negligible when $\eta < 1$. Indeed, for $\eta=3/4$ this correction is of about 60\%, 38\% and 16\% for $N=10^3$, $10^4$ and $10^6$, respectively. This potential corresponds to a 12$^{\mathrm{th}}$-power power-law trap, which can be obtained practically using $\ell=6$ Laguerre-Gaussian beams \cite{Mestre_2010}.

\begin{figure}[!h]
\centering
\includegraphics*[width=8.6cm,clip=true]{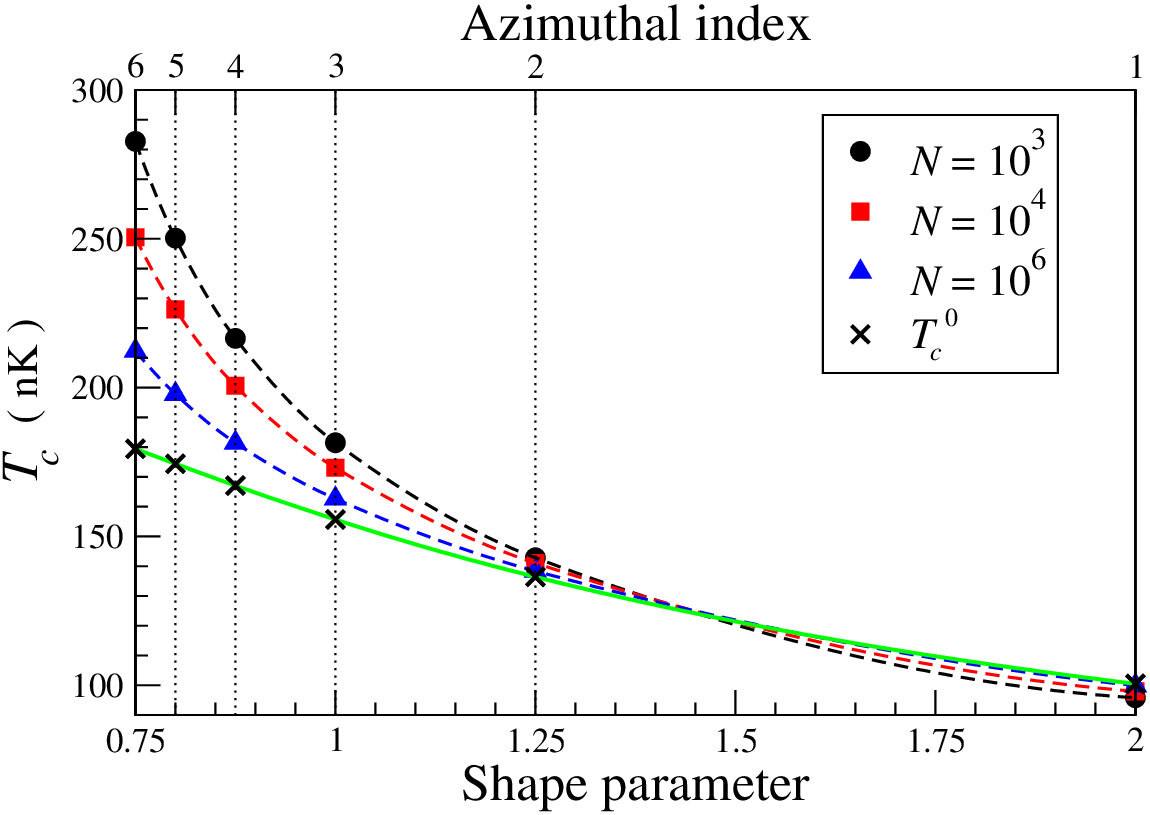}
\caption{(Color online) Condensation temperature in nK as a function of the shape parameter $\eta$ (lower scale) and of the azimuthal index $\ell$ (upper scale) for two Laguerre-Gaussian beams of power $P = 5\,$W and detuning $\delta = 10\,$THz. The solid green line with black crosses corresponds to the condensation temperature $T_c^0$ obtained in the thermodynamic limit. The dashed black line with circles gives the condensation temperature $T_c$ for $N = 10^3$ particles, the red dashed line with squares for $N=10^4$ and the blue dashed line with triangles for $N=10^6$. The volume of the trap has been chosen such that, for each atom number, the condensation temperature obtained in the thermodynamic limit for $\eta=2$ is 100\,nK (see \cite{Jaouadi_2010} for details).}
\label{Fig:Tc-l}
\end{figure}

Similarly, Fig. \ref{Fig:Tc-l} shows the variation of the condensation temperature with the shape parameter $\eta$ and the azimuthal index $\ell$. The temperature $T_c^0$ obtained in the thermodynamic limit is shown as a thick green solid line. It is identical for all atom numbers $N$ since we have chosen the volume of the trap such that the condensation temperature obtained in the thermodynamic limit is the same for all atom numbers (100\,nK with $\eta=2$, see \cite{Jaouadi_2010} for details). As already discussed in \cite{Jaouadi_2010}, the temperature $T_c^0$ increases with $\ell$ since $T_c^0 \propto N^{1/(\eta+1)}$ [see Eq. (\ref{Eq:Tc0})]. In the same figure, the condensation temperature $T_c$ taking into account finite size effects is shown as a dashed black line with circles for $N = 10^3$, a red dashed line with squares for $N = 10^4$ and a blue dashed line with triangles for $N = 10^6$. Compared to a harmonic trap of the same size, it can be concluded that much higher critical temperatures can be obtained in a power-law trap, and this especially for small condensation occupation numbers. This large increase results from the combination of two positive effects: first the effect of the potential shape on the temperature $T_c^0$ [Eq. (\ref{Eq:Tc0})] and second the finite size effect [Eq. (\ref{Eq:CompTc})]. Indeed, with the optical parameters chosen in Fig. \ref{Fig:Tc-l}, compared to $T_c^0 = 100\,$nK in the harmonic case ($\eta=2$), the condensation temperature for $\eta=3/4$ is about 283\,nK, 251\,nK and 212\,nK for $N=10^3$, $10^4$ and $10^6$, respectively.This corresponds to an increase of $T_c$ by about 183\%, 151\% and 112\%, respectively, when compared to the transition temperature found in a harmonic trap.

\section{Summary and concluding remarks}
\label{Sec:C}

In this paper, we have presented a detailed theoretical analysis of finite size effects on the transition temperature $T_c$ of an ideal Bose gas trapped in an inhomogeneous generic three-dimensional power-law potential. Using the local density approximation, we have derived an analytical expression of $T_c$, expressed in terms of a power series in $x_0=\varepsilon_0/k_BT_c$, where $\varepsilon_0$ denotes the zero-point energy of the trapping potential. We have also given numerical expressions truncated at second order for a specific set of power-law potentials.

More precisely, we have shown that in the case of a harmonic trap the usual lowest-order correction is inaccurate when the number particles $N$ is less than $10^5$. In this case, the next order has also to be taken into account, yielding to the relative correction
\begin{equation}
\frac{\Delta T_c}{T_c^0} \simeq
-0.7275\left(\frac{\omega_m}{\overline{\omega}}\right) N^{-\frac{1}{3}}
+ 1.098\left(\frac{\omega_m}{\overline{\omega}}\right)^{\!\!\frac{3}{2}} N^{-\frac{1}{2}}\,,
\label{Eq:HarmN-conclu}
\end{equation}
where $\omega_m$ and $\overline{\omega}$ denote the arithmetic and geometric means of the harmonic oscillator frequencies.

Beyond the specific case of the harmonic potential, we have also shown that, if the atoms are trapped in a power-law potential characterized by a shape parameter $\eta=3/2$, all finite size effects cancel out, and the critical temperature is simply equal to the one obtained in the thermodynamic limit, independently of the number of particles. In the case of a trap of cylindrical symmetry, this happens for example for $V(\mathbi{r}) \propto \rho^3 + |z|^3$, but also for $V(\mathbi{r}) \propto \rho^4 + |z|^2$.

In addition, we have shown that in a power-law potential characterized by a shape parameter $\eta  < 3/2$, the temperature shift due to finite size effects becomes positive. This shift may be relatively large for $\eta < 1$ with relatively small atom numbers ($N \leqslant 10^4$). The magnitude of this phenomenon was finally confirmed in the specific case of rubidium atoms confined by blue-detuned Laguerre-Gauss optical laser beams.

Finally, one should note that even though our approach gives the exact result for all types of power-law potentials within the limits imposed by the approximations used in our theoretical model, it is not able to capture the essence of critical phenomena taking place in a purely homogeneous system, due to the importance of long-range non-perturbative critical fluctuations in such systems.

\begin{acknowledgments}
The Institut des Sciences Mol\'eculaires d'Orsay (ISMO) is Unit\'e Mixte de Recherche de l'Universit\'e Paris-Sud 11 et du CNRS (UMR 8214). The authors would like to acknowledge financial support from Agence Nationale de la Recherche (Projects ImageFemto No. ANR-07-BLAN-0162-02 and AttoWave No. ANR-09-BLAN-0031-01) and from the R\'egion Ile-de-France (Programme SETCI).
\end{acknowledgments}


\end{document}